\documentclass[aps,prl,superscriptaddress]{revtex4-1}

\usepackage{amsmath}
\usepackage{graphicx}% Include figure files
\begin{document}

\title{Continuous wave lasing between Landau levels in graphene}

\author{Yongrui Wang}
\affiliation{Department of Physics and Astronomy, Texas A\&M
University, College Station, TX, 77843 USA}
\author{Mikhail Tokman}
\affiliation{Institute of Applied Physics, Russian Academy of Sciences}
\author{Alexey Belyanin}
\email{belyanin@tamu.edu}
\affiliation{Department of Physics and Astronomy, Texas A\&M
University, College Station, TX, 77843 USA}

\date{\today}

\begin{abstract}
We predict the general feasibility and demonstrate the specific design of the THz laser operating between Landau levels in graphene placed on a polar substrate in a magnetic field of order 1 T. Steady state operation under a continuous wave optical pumping is possible due to an interplay between Auger and surface-phonon mediated relaxation of carriers.  The scheme is scalable to other materials with massless Dirac fermions, for example surface states in 3D topological insulators such as Bi$_2$Se$_3$ or  Bi$_2$Te$_3$.  

\end{abstract}

%\keywords{}

\maketitle

\section{Introduction}

Free nonrelativistic electrons in a magnetic field behave as a system of harmonic oscillators, with selection rules allowing only the transitions between neighboring states with equal probabilities. Therefore they cannot be used as an active medium for lasers and masers. One way to get around this limitation is to accelerate electrons to high enough speeds that the relativistic effects become important. This leads to an anharmonicity in the electron spectrum and possibility of the maser action by accelerated electron beams, which has been so impressively implemented in vacuum electronic devices such as gyrotrons \cite{gyro}. Free carriers in semiconductors seem to offer a similar opportunity as the electron dispersion can show significant nonparabolicity above the bottom of the conduction band. Moreover, semiconductors offer a flexibility to grow heterostructures with different cyclotron transition energies which could be used for carrier injection into a given LL; see the proposal for a LL laser in the quantum Hall regime \cite{aoki1986}.  In practice, however, an ultrafast energy and momentum relaxation in semiconductors would quickly destroy population inversion between the Landau levels (LLs). As a result, there are no viable "solid state gyrotrons", although Landau level quantization does help with reducing scattering rate and improving  performance of quantum cascade lasers \cite{qcl} that operate through population inversion between quantum well subbands. 

Graphene seems to be an ideal material for the realization of LL lasers. Low energy excitations near the Dirac points in graphene have a linear conical spectrum which is obviously extremely nonparabolic. In a transverse magnetic field the 2D conical spectrum splits into a series of non-equidistant LLs with energies scaling as a square root of the magnetic field and the principal quantum number. It was suggested in \cite{morimoto2008} that the optical pumping to an arbitrary  excited state $n \geq 1$ will lead to electrons cascading down the LLs preferentially emitting photons, which would potentially lead to the EM field amplification on any of these  downward transitions. Unfortunately, the proposal  \cite{morimoto2008} assumed that the radiative transitions are the fastest ones in graphene. It did not include most important nonradiative relaxation channels and did not attempt to calculate actual LL populations. In particular, it turns out that the Auger relaxation is a very powerful relaxation mechanism for Dirac electrons in a magnetized graphene that proceeds much faster than radiative transitions and washes out any population inversion over the time scale of few ps; see below and also recent theoretical calculations of the Auger relaxation rate \cite{knorr} and experimental measurements in \cite{helm2014}.  A recently proposed, more sophisticated pumping scheme \cite{malic2014} takes into account Auger relaxation processes and still leads to only a transient population inversion existing over a ps timescale. 

Here we propose what we believe is a viable inter-LL laser scheme for graphene that takes into account all relevant relaxation processes and in fact utilizes them to reach a steady-state population inversion, vital for any viable laser. Our scheme is transferable to thin ($\lambda \gg \Delta z \geq 5$ nm) films of 3D topological insulators such as Bi$_2$Se$_3$ where the Landau levels associated with massless metallic surface states  \cite{LLs2, LLs3} should demonstrate similar coupling to the EM field despite different chirality \cite{yao2014}.  
Not that our scheme provides the population inversion {\it in a steady state}, i.e. under a continuous-wave pumping, in contrast to previous proposals, with or without the magnetic field, that could provide only a transient gain during a picosecond time interval  \cite{morimoto2008,otsuji,malic2014}. 

We solve kinetic and density matrix equations coupled with Maxwell's equations to calculate populations, gain and laser threshold conditions as a function of the optical pumping power. The calculation  details are in the sections below. Here we present a general idea of the laser scheme. It is illustrated in Fig.~1. It shows one specific implementation of the scheme with an optical pumping originated from level $ n = -2$ to  obtain maximum population inversion between levels -1 and -2.  However, the scheme can me implemented for any pair of LLs $(-n, -n-1)$ as long as level $-n$ stays deep enough below the Fermi level. The lasing wavelength can be from sub-THz to the mid-infrared range, depending on the value of $n$, the magnetic field, and the substrate used. 
\begin{figure}[htbp]
\includegraphics[width=0.5\textwidth]{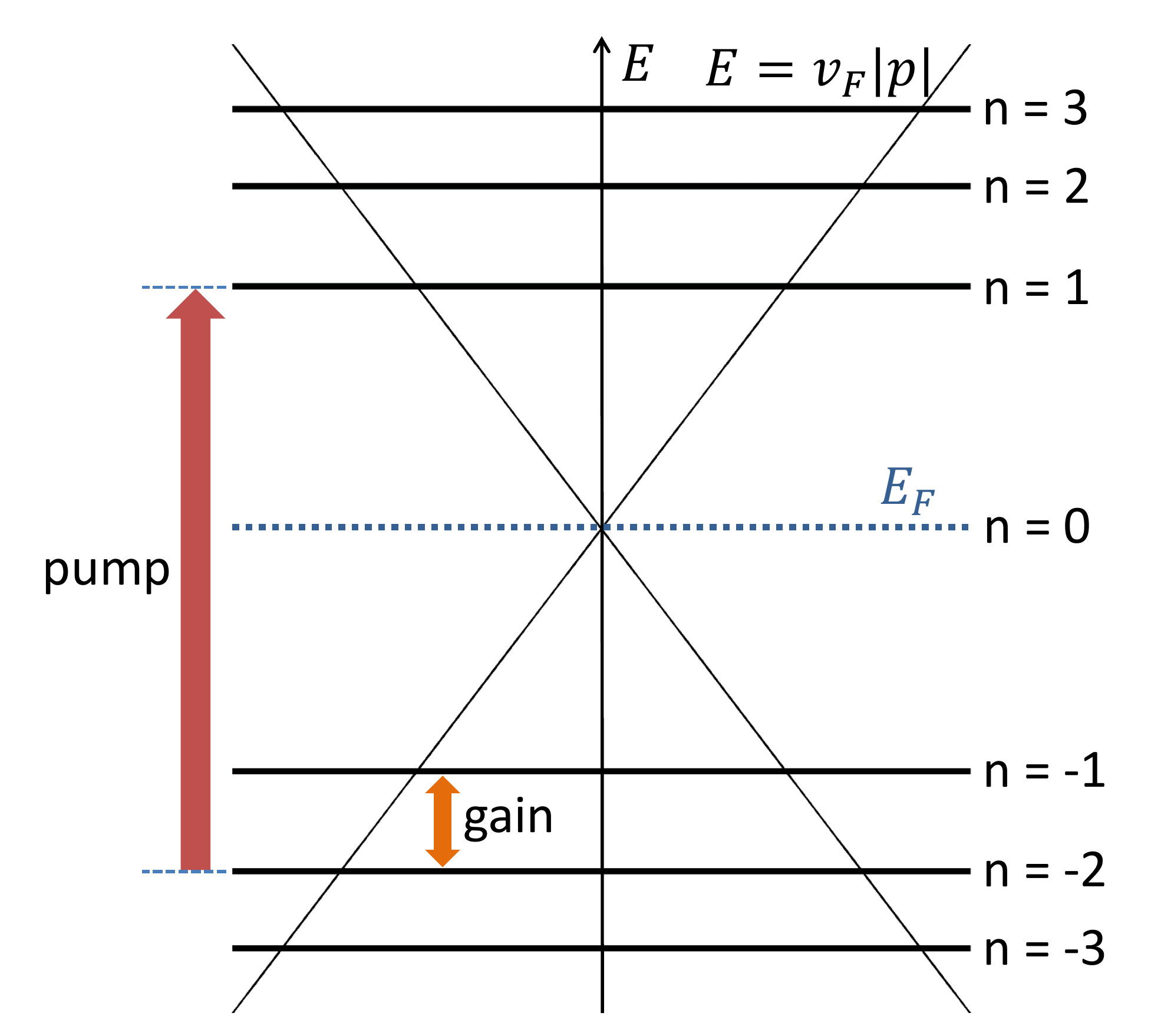}
\caption{The scheme to obtain population inversion between the electron states below the Fermi level by using a continuous-wave optical pumping.\label{Fig:pumping_scheme}}
\end{figure}

The general idea is as follows. In the equilibrium (without pumping) labels $n = -1$ and $-2$ are fully occupied to degeneracy surface density $N_{s}=g_s g_v/2\pi {l_c}^2$, where spin and valley degeneracy factors are $g_s = 2$ and $g_v = 2$ for graphene, and $l_c=\sqrt{c\hbar/eB}$ is the magnetic length. The Fermi level is placed at the Dirac point in the figure, assuming intrinsic graphene. However, this can be changed, as long as level $n = -1$ is fully occupied in equilibrium. An optical pumping resonant to the transition $-2 \rightarrow 1$ moves part of the carriers up from level $ n = -2$ creating a population inversion between a fully occupied level $n = -1$ 
and level $-2$. In order for this population inversion to exist in a steady state, i.e. under a continuous-wave optical pumping, the relaxation of carriers back to lower laser state $-2$ should be slower than the relaxation rate to the upper laser state $-1$. Unfortunately, the Auger mechanism does not satisfy this criterion. Our simulations show that there is no population inversion in the steady state, no matter how strong the optical pumping power is.  This is because an increase in the  depopulation rate of level $-2$ by an optical pumping is compensated by an increase in the Auger scattering rate to level 2, primarily through the scattering of electrons from states in levels 1 and $-1$ to states in levels 2  and $-2$, respectively.  
In order to overcome this obstacle, the magnetic field needs to be tuned in order to bring the transitions $ 1 \rightarrow 0$ and $0 \rightarrow -1$ (of the same energy) in resonance with an LO phonon energy. This will greatly increase the rate of electron relaxation from excited states to the upper laser state $ n = -1$ through LO phonon emission, whereas the transitions to state $-2$ will be out of resonance and not affected much. 

The LO phonon energy in graphene is close to 200 meV, which would require a magnetic field of almost 30 T to bring the transition frequency $\omega_{10}$ close to $\omega_{LO}$. In order to reduce the required magnetic field one can utilize the scattering on bulk, surface, or interface optical phonons of the substrate, and choose the substrate with a lower optical phonon energy, for example a polar semiconductor such as GaAs or InGaAs \cite{stoberl2008}. For definiteness, below we assume the substrate to be GaAs, which leads to the surface optical (SO) phonon energy of $\hbar \omega_{SO} = 36$ meV \cite{dubois1982}.  This is equal to $\omega_{01} = \omega_c = \sqrt{2} v_f / l_c$ in a magnetic field of 1 T. The laser transition wavelength would be then around 82 $\mu$m, i.e. around 3 THz, which is the range where there is a shortage of laser sources. We will also assume the optical pumping between levels $-2$ and 1, although the pumping resonant to the transition from $-2$ to 3 would be equally efficient and lead to a similar value for the gain. Moreover, the transition frequency for the latter transition in a magnetic field of 1 T would correspond to a CO$_2$ laser wavelength around 10 $\mu$m, which could be more convenient than the 14-$\mu$m wavelength corresponding to the transition $-2 \rightarrow 1$. Of course all energies can be changed as needed by choosing different substrates or different LLs for the lower laser state, for example $ n = -3$ instead of $-2$.

\section{Electron states and optical transitions between the Landau levels in graphene}

For completeness, we give a brief summary of the electron states and optical transitions between the LLs in graphene, since this information is extensively used below. They have been calculated many times before and observed both in monolayer and multilayer samples \cite{orlita}.

 Neglecting intervalley scattering, we will only need electron states in one of the two equivalent $K,K'$ valleys, for example the $\vec{K}$ valley. Without a magnetic field, the low-energy Hamiltonian in the vicinity of the $\vec{K}$ Dirac point  is given by \cite{Goerbig_graphene_magnetic_RevModPhys_2011}
\begin{eqnarray}
H = v_{F} \vec{\sigma} \cdot \hat{\vec{p}}
 = v_F \begin{pmatrix}
0 & \hat{p}_x-i \hat{p}_y \\
\hat{p}_x+i \hat{p}_y & 0
 \end{pmatrix}~,
\end{eqnarray}
where $v_F = 10^8$ cm/s. In the presence of a transverse magnetic field or any EM field described by the vector-potential $A$, we replace $\hat{\vec{p}}$ with $\hat{\vec{\Pi}}$ = $\hat{\vec{p}} + e \vec{A}/c$. For a magnetic field in the $+z$ direction,  we can write $\vec{A}$ = $(0,~Bx,~0)$ in the Landau gauge, then the eigenfunctions are expressed as \cite{Ando_graphene_2002}
\begin{eqnarray}
\label{Eq:wave_func}
F^K_{nk}(\vec{r}) = \frac{1}{\sqrt{L}} e^{i k y} \Phi_n(k, x) ~,
\end{eqnarray}
with
\begin{eqnarray}
\Phi_n(k, x) = C_n
\begin{pmatrix}
\mathrm{sgn}(n)i^{|n|-1}\phi_{|n|-1}(x + l_c^2 k) \\
i^{|n|} \phi_{|n|}(x + l_c^2 k)
\end{pmatrix}~,
\end{eqnarray}
where  $C_n$ = 1 when n = 0, and $C_n$ = $1/\sqrt{2}$ when n $\neq$ 0; $\mathrm{sgn}(x)$ = 1, 0, -1 for x $>$ 0, x = 0, x $<$ 0 respectively. The function $\phi_{|n|}(x)$ has the same form as the eigenfunction in the massive electron case:
\begin{eqnarray}
\phi_{|n|}(x) = \frac{1}{\sqrt{2^{|n|}|n|!\sqrt{\pi}l_c}} \mathrm{exp} \left[ -\frac{1}{2} \left( \frac{x}{l_c}\right)^2 \right] H_{|n|}\left( \frac{x}{l_c}\right) ~,
\end{eqnarray}
where $H_{|n|}(x)$ is the Hermite polynomial. The corresponding eigenenergy is $\epsilon_n$ = $\mathrm{sgn}(n) \hbar \omega_c \sqrt{|n|}$, with $\omega_c$ = $\sqrt{2} v_f / l_c$.

In this manuscript, an electron state will be labeled by $| n, k, s, \xi \rangle$, where s = \{$\uparrow$, $\downarrow$\} denotes spin,  $\xi$ = \{$\vec{K}$, $\vec{K'}$\} denotes valley; $k$, $s$, $\xi$ are degenerate quantum numbers, and the total degeneracy density of a Landau level $n$ is $2/\pi l_c^2$.

 The interaction Hamiltonian for an optical field with an in-plane polarization can be written as
 \begin{eqnarray}
 \hat{H}^{op}_{int} = v_F \frac{e}{c} \vec{\sigma} \cdot \vec{A}^{op}~,
 \end{eqnarray}
where $\vec{A}^{op}$ is the vector potential of the optical field, which is related to the electric field by $\vec{E}^{op}$ = $(-1/c) \partial\vec{A}^{op} / \partial t$. If we define two circular polarization vectors, $\hat{l}_\oplus$ = $(\hat{x}+i\hat{y})/\sqrt{2}$ and $\hat{l}_\ominus$ = $(\hat{x}-i\hat{y})/\sqrt{2}$, the vector potential of a single frequency optical field can be written as
\begin{eqnarray}
 \vec{A}^{op} = \frac{1}{2} \left( A_\oplus \hat{l}_\oplus + A_\ominus \hat{l}_\ominus \right) e^{-i\omega t} + c.c. ~.
 \end{eqnarray}
 Plugging this expression into the Schr\"{o}dinger equation and using the rotating wave approximation, we get the same selection rules as in \cite{PhysRevB.75.155430}: Transitions between $n_1$ and $n_2$ ($n_2$ $>$ $n_1$) are coupled by photons with $\hat{l}_\oplus$ polarization if $|n_2|$ = $|n_1|$ + 1, and with $\hat{l}_\ominus$ polarization if $|n_2|$ = $|n_1|$ - 1. 
 
 By expressing $\vec{A}^{op}$  through $\vec{E}^{op}$ in $\hat{H}^{op}_{int}$, we can get the magnitude of the dipole moment for a resonant transition between Landau levels $n_1$ and $n_2$:  
 \begin{eqnarray}
 |\mu_{n_1 n_2}| = \sqrt{2} C_{n_1} C_{n_2} e v_F / \omega~.
 \end{eqnarray}
 The two dimensional linear optical susceptibility near the resonance to the transition between $n_1$ and $n_2$ ($n_1$ $<$ $n_2$) is
 \begin{eqnarray}
 \chi_{n_1 n_2} = \frac{2}{\pi l_c^2} \frac{|\mu_{n_1 n_2}|^2 (f_{n_2}-f_{n_1})}{\hbar\omega - (\epsilon_{n_2}-\epsilon_{n_1}) + i \hbar/T_2 } ~,
 \end{eqnarray}
 where $T_2$ is the dephasing time. The optical transition rate between $n_1$ and $n_2$ ($n_1$ $<$ $n_2$) is 
\begin{eqnarray}
\Gamma^{op}_{n_1 n_2} =   \frac{1}{2} \left| \frac{\mu_{n_1 n_2} E^{op}}{\hbar} \right|^2 \frac{1/T_2}{(1/T_2)^2 + ((\epsilon_{n_2}-\epsilon_{n_1})/\hbar - \omega)^2}~.
\end{eqnarray}

\section{Laser threshold condition}

To determine the threshold condition for a LL graphene laser we consider the simplest geometry resembling a quantum-well vertical cavity laser, in which an active layer consisting of one or several graphene monolayers on a polar substrate is located between the two mirrors of given reflection factors $r_{1,2}$; see Fig.~2. We will assume that there are  two media with dielectric constants $\kappa_1$ and $\kappa_2$  from both sides of the active layer.  We will also assume for simplicity that the thickness of an active layer  is much smaller than the wavelength of the THz laser field. For a field of amplitude $E_i$ incident on the graphene layer, the amplitudes of reflected and transmitted waves $E_r$ and $E_t$ can be related using the Maxwell's equations with proper boundary conditions as

\begin{figure}[htbp]
\includegraphics[width=0.5\textwidth]{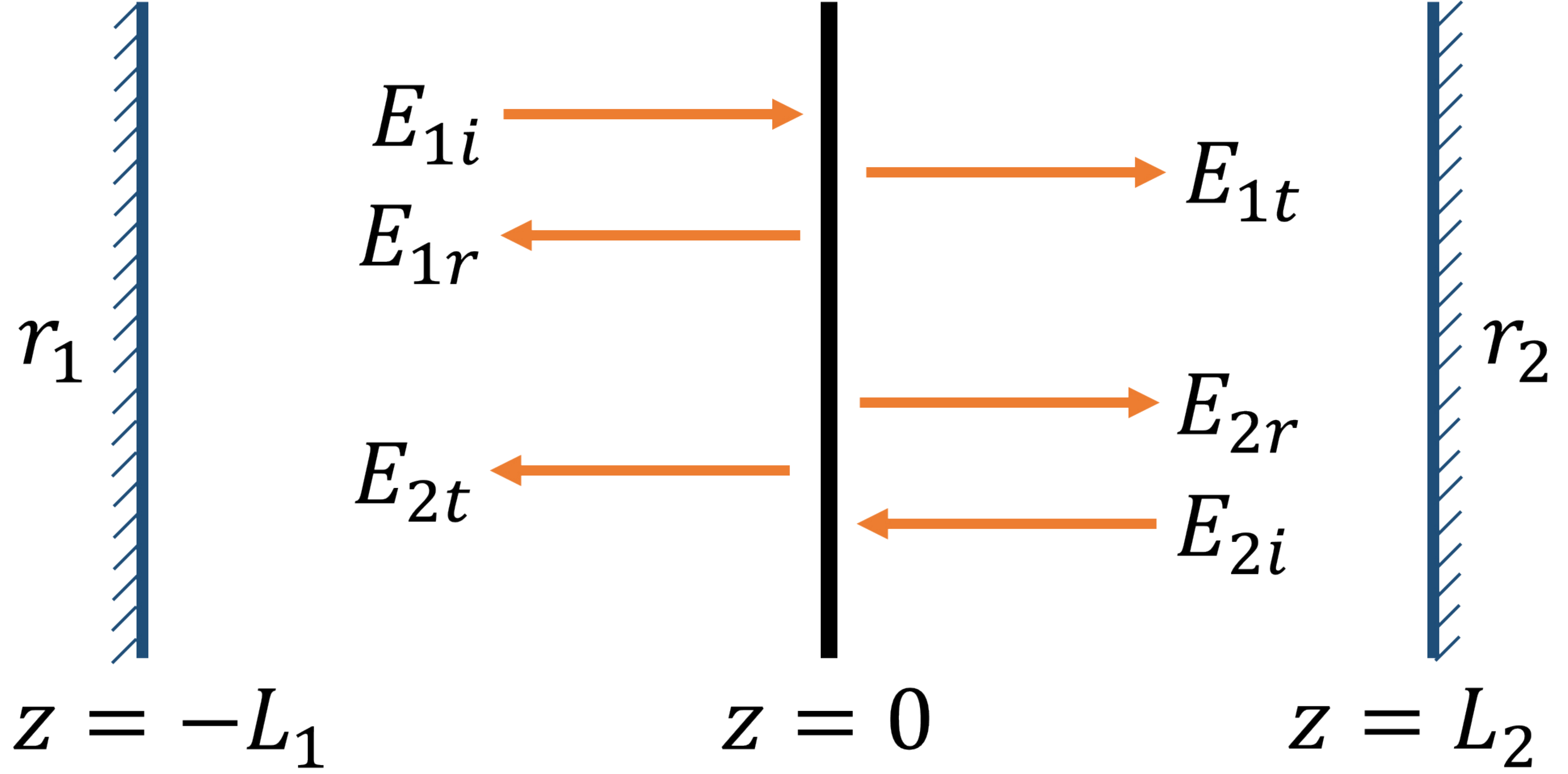}
\caption{A design of graphene laser.\label{Fig:cavity}}
\end{figure}

%\begin{figure}[htbp]
%\includegraphics[width=0.5\textwidth]{gain2D.pdf}
%\caption{Optical field incident on graphene layers.\label{Fig:gain_2D}}
%\end{figure}

\begin{eqnarray}
\label{boundary}
&E_t = \dfrac{2}{1+\sqrt{\frac{\kappa_2}{\kappa_1}} - \frac{4\pi i\omega}{\sqrt{\kappa_1}c} \chi } E_i ~, \nonumber  \\
&E_r = \left(\dfrac{2}{1+\sqrt{\frac{\kappa_2}{\kappa_1}} - \frac{4\pi i\omega}{\sqrt{\kappa_1}c} \chi } -1 \right) E_i ~.
\end{eqnarray}
To make equations even simpler, we will take  $\kappa_1$ = $\kappa_2$ =$\kappa$. It is straightforward to include more complex cavity structures if needed for a particular design. 

The fields also need to satisfy the boundary conditions at the mirrors:
\begin{eqnarray}
\label{mirror}
E_{1i} e^{-i k L_1} = r_1 (E_{1r} + E_{2t}) e^{i k L_1} ~, \nonumber \\
E_{2i} e^{-i k L_2} = r_2 (E_{1t} + E_{2r}) e^{i k L_2} ~.
\end{eqnarray}
From the boundary conditions Eqs.~(\ref{boundary}) and (\ref{mirror}), the condition to have stable nonzero optical fields inside the cavity is
\begin{eqnarray}
\label{Eq:chi_requirement}
 - \dfrac{2\pi i\omega}{\sqrt{\kappa}c} \chi = \dfrac{r_1 r_2 - e^{-2 i k (L_1 + L_2)}}{r_1 r_2 + r_1 e^{-2 i k L_2} + r_2 e^{-2 i k L_1} + e^{-2 i k (L_1 + L_2)}}~.
\end{eqnarray}
To get the threshold, we assume that the optical fields are in resonance with respective transitions and the lengths $L_1$ and $L_2$ are adjusted so that every term in the denominator has the same sign. Then the minimum required imaginary part of the susceptibility in the active layer can be found from the real part of Eq.~(\ref{Eq:chi_requirement}): 
\begin{eqnarray}
\label{Eq:gain_requirement}
- \dfrac{2\pi \omega}{\sqrt{\kappa}c} {\mathrm Im}[\chi] = \dfrac{1- |r_1 r_2|}{1 + |r_1| + |r_2| + |r_1 r_2|}~.
\end{eqnarray}
We will discuss the feasibility of reaching the lasing threshold below, after calculating the rates of scattering processes, the non-equilibrium populations of the LLs, and the resulting graphene susceptibility at the laser transition in the presence of an optical pumping.   

The carriers excited by an optical pumping relax through a variety of scattering processes. The steady state populations are determined by a balance between relaxation and the continuous wave pumping. In the next two sections we give a detailed description of most important processes that determine the redistribution of populations and the resulting steady-state gain. 

\section{Auger processes}

A strong magnetic field suppresses scattering processes due to energy quantization and reduction in the phase space available for scattered carriers. However, Auger processes remain very efficient: due to the symmetry between electron and hole LLs there is always resonance for scattering of carriers from $(0,0)$ LLs into $(1,-1)$ states and for all other combinations allowed by the energy conservation: $(1,-1) \leftrightarrow (2,-2)$, $(0,0) \leftrightarrow (2,-2)$, $(-1,1) \leftrightarrow (2,-2)$, etc.; see Fig.~3. Recently the Auger relaxation rates were measured to be in a few ps range in pump-probe experiments \cite{helm2014}, which agrees with our simulations. Below we outline the general derivation of the Auger scattering rate and then apply it to our problem. 

\begin{figure}[htbp]
\includegraphics[width=0.5\textwidth]{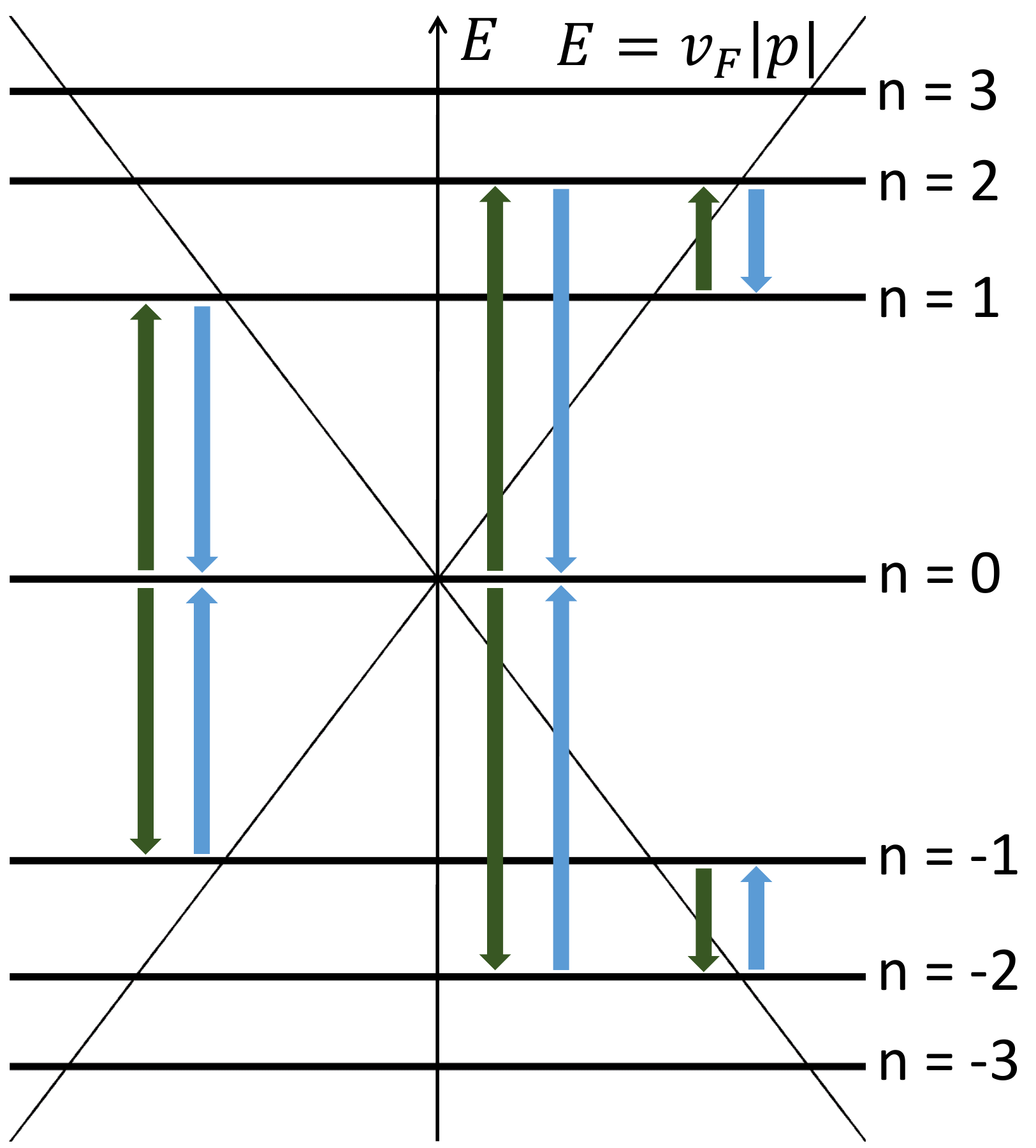}
\caption{Examples of the Auger scattering processes between $n = 0,\pm 1, \pm 2$ LLs. .\label{Fig:auger}}
\end{figure}

\subsection{General formulas}
Auger processes are mediated by the Coulomb interaction between carriers. The general Coulomb interaction Hamiltonian for electrons can be written as \cite{nolting2009manybody}
\begin{eqnarray}
V_C = \frac{1}{2} \sum_{\alpha\beta\gamma\delta} V_{\alpha\beta\gamma\delta} a_\alpha^\dagger a_\beta^\dagger a_\delta a_\gamma ~,
\end{eqnarray}
where
\begin{eqnarray}
V_{\alpha\beta\gamma\delta} = \langle \alpha(1)| \langle\beta(2)| V_{Coul}(\vec{r}_1 - \vec{r}_2) |\gamma(1)\rangle |\delta(2)\rangle ~.
\end{eqnarray}
In order to simplify this expression and include the effect of screening, we expand $V_{Coul}(\vec{r}_1 - \vec{r}_2)$ in  Fourier series 
\begin{eqnarray}
V_{Coul}(\vec{r}_1 - \vec{r}_2) = \sum_{\vec{q}} V_{\vec{q}}~ e^{i \vec{q} \cdot (\vec{r}_1 - \vec{r}_2)}~,
\end{eqnarray}
where $V_{\vec{q}}$ = $2\pi e^2/\kappa_0 A q$ for a 2-dimensional case. Using this expression, we get
\begin{eqnarray}
\label{Eq:VCoulomb_calculated}
V_{\alpha\beta\gamma\delta} = \sum_{\vec{q}} V_{\vec{q}}  \langle \alpha(1)| e^{i \vec{q}\cdot\vec{r}_1} |\gamma(1)\rangle \langle\beta(2)| e^{-i \vec{q}\cdot\vec{r}_2} |\delta(2)\rangle ~.
\end{eqnarray}
To include screening, we replace $V_{\vec{q}}$ with $V_s(\vec{q},\omega)$ = $V_{\vec{q}}/\epsilon(\vec{q},\omega)$, where the dielectric function $\epsilon(\vec{q},\omega)$ in the random phase approximation is given by the Lindhard formula
\begin{eqnarray}
\label{Eq:screening_Lindhard}
\epsilon(\vec{q}, \omega) = 1 - V_{\vec{q}} \Pi^0 (\vec{q}, \omega)~,
\end{eqnarray}
and the polarizability $\Pi^0 (\vec{q}, \omega)$ is written as
\begin{eqnarray}
\label{Eq:polarizability}
\Pi^0 (\vec{q}, \omega) = \sum_{\alpha\beta} \frac{f_\alpha - f_\beta}{\epsilon_\alpha - \epsilon_\beta + \hbar\omega + i \delta} |F_{\alpha\beta}(\vec{q})|^2 ~,
\end{eqnarray}
with the form factor $F_{\alpha\beta}(\vec{q})$ = $\langle\alpha|e^{i\vec{q}\cdot\vec{r}}|\beta\rangle$. The value of $\omega$ is determined by $\hbar\omega$ = $E_\gamma - E_\alpha$ in $V_{\alpha\beta\gamma\delta}$ \cite{omega_screening_PhysRevB.88.035430}. The rate of the Auger scattering from state $|a,b\rangle$ to state $|c,d\rangle$ is calculated from the Fermi's golden rule; it is symmetric with respect to the initial and final states: 
\begin{eqnarray}
\label{Eq:Auger_rate}
\Gamma_{ab\leftrightarrow cd} = \frac{2\pi}{\hbar} \left| \langle c d| V_C | a b \rangle \right|^2 \delta(E_c + E_d - E_a - E_b) ~,
\end{eqnarray}
where the matrix element is
\begin{eqnarray}
\label{matrix}
\langle c d| V_C | a b \rangle = \frac{1}{2} \left( V_{cdab} - V_{dcab} + V_{dcba} - V_{cdba} \right)~.
\end{eqnarray}
So, there are essentially four terms because electrons are indistinguishable particles. The state $|a,b\rangle$ in Eq.~(\ref{matrix}) only means that both $|a\rangle$ and $|b\rangle$ are occupied, instead of specifying that electron 1 is in $|a\rangle$, and electron 2 is in $|b\rangle$. One can find mistakes in the literature with some of the terms missing. 
% one example is in reference \cite{Auger_wrong_PhysRevB.37.2578}.

\subsection{Auger scattering between Landau levels in graphene}

For graphene in a transverse magnetic field, an electron state can be written as $|\alpha\rangle$ = $| n_\alpha, k_\alpha, s_\alpha, \xi_\alpha \rangle$, with notations explained in Sec.~II. We will address the screening effect first. The form factor in Eq.\,(\ref{Eq:polarizability}) can be evaluated to be
\begin{eqnarray}
F_{\alpha\beta}(\vec{q}) = \langle\alpha|e^{i\vec{q}\cdot\vec{r}}|\beta\rangle = \delta_{s_\alpha,s_\beta} \delta_{\xi_\alpha,\xi_\beta} \delta_{q_y, k_\alpha-k_\beta} e^{-i q_x l_c^2 k_\beta} G_{n_\alpha n_\beta}(q_y, q_x)~.
\end{eqnarray} 
So the polarizability becomes
\begin{eqnarray}
\label{Eq:pol_grapheneLL}
\Pi^0 (\vec{q}, \omega) = \frac{2 A}{\pi l_c^2} \sum_{n_\alpha n_\beta} \frac{f_{n_\alpha} - f_{n_\beta}}{\epsilon_{n_\alpha} - \epsilon_{n_\beta} + \hbar\omega + i\delta} |G_{n_\alpha n_\beta}(q_y, q_x)|^2~.
\end{eqnarray}
One can check that $|G_{n_\alpha n_\beta}(q_y, q_x)|$ only depends on the magnitude of $\vec{q}$, so $\Pi^0(\vec{q},\omega)$ and $V_s(\vec{q},\omega)$ are only functions of $q$ = $|\vec{q}|$. We calculated the form factor numerically and checked that it agreed with the analytical expression in \cite{Goerbig_graphene_magnetic_RevModPhys_2011}.

Then, using Eq.\,(\ref{Eq:VCoulomb_calculated}) and Eq.\,(\ref{Eq:Mfi_calculated}), the Coulomb matrix element is
\begin{align}
 V_{abcd} &= \delta_{s_a,s_c} \delta_{s_b,s_d} \delta_{\xi_a,\xi_c} \delta_{\xi_b,\xi_d} \nonumber \\
&\times \sum_{\vec{q}} V_s(\vec{q},\omega) \delta_{q_y, k_a-k_c} e^{-i q_x l_c^2 k_c} G_{n_a n_c}(q_y, q_x) \delta_{-q_y, k_b-k_d} e^{i q_x l_c^2 k_d} G_{n_b n_d}(-q_y, -q_x) \nonumber \\
& = \delta_{s_a,s_c} \delta_{s_b,s_d} \delta_{\xi_a,\xi_c} \delta_{\xi_b,\xi_d} \delta_{k_a + k_b, k_c + k_d} \nonumber \\
&\times \sum_{q_x} \left.V_s\left(\vec{q},\omega\right) e^{-i q_x l_c^2 (k_c - k_d)} G_{n_a n_c}(q_y, q_x) G_{n_b n_d}(-q_y, -q_x)\right|_{q_y = k_a-k_c} ~.
\end{align}
For a fixed $k_a$, this matrix element decays quickly when $k_c-k_a$ is large, since $G_{n_a n_c}(q_y, q_x)$\,$\propto$\,$\exp(-(q l_c)^2/4)$ when $q$ is large \cite{Goerbig_graphene_magnetic_RevModPhys_2011}. If $k_c$ is bounded, then $k_b$ is bounded too, otherwise the term $e^{-i q_x l_c^2 (k_c - k_d)}$ would oscillate too fast with $q_x$, which essentially makes the summation over $q_x$ to vanish.

 The Auger scattering rate between two pairs of Landau levels $(n_a, n_b)$ and $(n_c, n_d)$ is
\begin{eqnarray}
\label{rate} 
\Gamma_{n_a n_b \leftrightarrow n_c n_d} = \frac{1}{2 A /\pi l_c^2} \sum_{\xi_a,k_a,s_a} \sum_{\xi_b,k_b,s_b} \sum_{\xi_c,k_c,s_c} \sum_{\xi_d,k_d,s_d} \Gamma_{ab\leftrightarrow cd} \left( \times \frac{1}{2} \mbox{ if } n_a = n_b \mbox{ or } n_c = n_d \right)~,
\end{eqnarray}
where the factor 1/2 in the parenthesis is because of the double counting the initial or final states. In Eq.~(\ref{rate}), one of the summations can be dropped immediately, since the result from the other three summations will be independent of the forth set of quantum numbers. This summation will give exactly the degeneracy $2 A /\pi l_c^2$, so it will cancel with the pre-factor. One summation of $k$ can also be eliminated due to the conservation of momentum $k_a+k_b$ = $k_c+k_d$. As the energy is fully quantized, we will replace the $\delta$ function in Eq.\,(\ref{Eq:Auger_rate}) with a Lorentzian of line width which can be attributed to impurity scattering \cite{Goerbig_graphene_magnetic_RevModPhys_2011}.

\section{Phonon scattering}

\subsection{General formulas}

The interaction Hamiltonian between phonons and electrons can generally be written as
\begin{eqnarray}
H_{int}^{ph} = \sum_{\vec{k},~\vec{q}} F(q) c_{\vec{k}+\vec{q}}^\dagger c_{\vec{k}} ( b_{\vec{q}} + b_{-\vec{q}}^\dagger )~,
\end{eqnarray}
where $c$ and $c^\dagger$ are annihilation and creation operators for electrons, $b$ and $b^\dagger$ are annihilation and creation operators for phonons, and $F(q)$ is defined below. 
Using Fermi's golden rule, the scattering rate from an initial electronic state $|\varphi_i\rangle$ to a final state $|\varphi_f\rangle$ is
\begin{gather}
\Gamma^{ph}_{i \to f} = \frac{2\pi}{\hbar} \sum_{\vec{q}} (n_q+1) |F(q)|^2 |M_{fi}(\vec{q})|^2 \delta(\epsilon_f+E^{ph}_q - \epsilon_i) \nonumber \\
+ \frac{2\pi}{\hbar} \sum_{\vec{q}} n_q |F(q)|^2 |M_{fi}(\vec{q})|^2 \delta(\epsilon_f - E^{ph}_q - \epsilon_i) ~,
\end{gather}
where the first term is for the phonon emission, the second term is for the phonon absorption, and the matrix element is given by
\begin{gather}
M_{fi}(\vec{q}) = \langle \varphi_f | \sum_{\vec{k}} c_{\vec{k}+\vec{q}}^\dagger c_{\vec{k}} | \varphi_i \rangle \nonumber \\
= \int{d \vec{r} \varphi_f^\dagger(\vec{r}) e^{i\vec{q}\cdot\vec{r}} \varphi_i(\vec{r}) }
\end{gather}
Using the wave functions in Eq.\,(\ref{Eq:wave_func}), the matrix element $M_{fi}(\vec{q})$ for $|i\rangle$ = $|n_i, k_i, s, \xi \rangle$ and $|f\rangle$ = $|n_f, k_f, s, \xi \rangle$ is calculated to be
\begin{eqnarray}
\label{Eq:Mfi_calculated}
\delta_{q_y, k_f-k_i} e^{-i q_x l_c^2 k_i} G_{n_f n_i}(q_y, q_x) ~,
\end{eqnarray}
where we have defined
\begin{eqnarray}
\label{Eq:G_factor}
G_{n_1 n_2}(q_y, q_x) \equiv  \int d x \Phi_{n_1}^\dagger (q_y, x) e^{i q_x x} \Phi_{n_2} (0, x) ~.
\end{eqnarray}
The averaged scattering rate from an initial Landau level $n_i$ to a a final Landau level $n_f$ is 
\begin{align}
\label{Eq:phonon_rate}
\Gamma_{n_i \to n_f}^{ph} &= \sum_{k_f} \Gamma_{i \to f}^{ph} \nonumber \\
&= \frac{2\pi}{\hbar} \sum_{\vec{q}} (n_q+1) |F(q)|^2 |G_{n_f n_i}(q_y, q_x)|^2 \delta(\epsilon_f+E^{ph}_q - \epsilon_i) \nonumber \\
&+ \frac{2\pi}{\hbar} \sum_{\vec{q}} n_q |F(q)|^2 |G_{n_f n_i}(q_y, q_x)|^2 \delta(\epsilon_f - E^{ph}_q - \epsilon_i) ~.
\end{align}

We include the dynamic screening effect by carriers in graphene in the phonon scattering processes. This can be done by replacing $F(q)$ with $F_s(q,\omega)$ = $F(q)/\epsilon(q,\omega)$, where the dielectric function $\epsilon(q,\omega)$ is given in Eq.\,(\ref{Eq:screening_Lindhard}) and Eq.\,(\ref{Eq:pol_grapheneLL}). We will only consider the case of  low enough temperatures, when phonon absorption is unimportant and only phonon emission processes contribute to the scattering rate. At room temperature this is still a reasonable approximation; it can be easily dropped if a greater accuracy is needed.

\subsection{LA phonon scattering}

For longitudinal acoustic (LA) phonon scattering, the expression for $F(q)$ is \cite{LAphonon_graphene}
\begin{eqnarray}
F_{LA}(q) = -\sqrt{\frac{\hbar}{2 \rho A v_s}} D \sqrt{q} ~,
\end{eqnarray}
where $\rho = 7.6 \times 10^{-8}$ g/cm$^2$ is the area mass density of graphene, $v_s = 2\times 10^6$ cm/s is the sound velocity, and $D$ is in the 10 - 50 eV range. Also, the energy of a LA phonon is $E^{LA}_q = \hbar v_s q$. Plugging these expressions into Eq.\,(\ref{Eq:phonon_rate}), we get the scattering rate by LA phonons:
\begin{eqnarray}
\label{Eq:LArate}
\Gamma^{LA}_{n_i \to n_f} = \frac{D^2 q_0^2}{4\pi \rho \hbar v_s^2} \int_0^{2\pi} d\theta |G_{n_f n_i}(q_0 \sin\theta, q_0\cos\theta)|^2 ~,
\end{eqnarray}
where $q_0 = (\epsilon_{n_i} - \epsilon_{n_f})/\hbar v_s$. The coefficient is of the order of $10^{14}$ s$^{-1}$ for B $\sim$ 1\,T. However, the integrand in Eq.\,(\ref{Eq:LArate}) is roughly of the order of $\exp[-(q_0 l_c)^2/2]$, which is extremely small, since $q_0 l_c \sim \omega_c l_c/v_s = \sqrt{2}v_F/v_s = 50 \sqrt{2}$. So, the LA phonon scattering does not contribute significantly to electronic transitions between Landau levels due to a large ratio $v_f / v_s \gg 1$.

\subsection{Surface optical phonon scattering}

Since we want to use phonon scattering to our advantage in order to facilitate electron relaxation to the upper laser state, we consider graphene on a polar substrate or sandwiched between two substrates.  In this case, the electrons in graphene can couple to the surface or interface modes of optical phonons \cite{dubois1982, Ando_SOphonon_1989}, which we will call the surface optical (SO) phonons for brevity. If the two substrates on both sides of the graphene layer are the same, the SO phonon energy is equal to the longitudinal optical (LO) phonon energy of the substrate \cite{Ando_SOphonon_1989}. If there is vacuum on one side, the SO phonon energy is slightly shifted from the LO phonon energy \cite{dubois1982}. We will assume the former case for definiteness, but note that it would be straightforward to calculate interface optical phonon modes for an arbitrarily complex structure. If we assume that the graphene layer does not affect the SO phonon modes, then the expression of $F(q)$ can be written as 
\begin{eqnarray}
F_{SO}(q) = \left. \left[ \frac{2\pi e^2\hbar \omega_{SO}}{A} \left( \frac{1}{\kappa_\infty^{sub}} - \frac{1}{\kappa_0^{sub}} \right) \right] \middle/ {\sqrt{2 q}} \right. ~,
\end{eqnarray}
where $A$ is the area of graphene, $\kappa_0^{sub}$ ($\kappa_\infty^{sub}$) is the low (high) frequency dielectric constant of the substrate, and $\hbar\omega_{SO}$ is the energy of the surface optical phonon. Since it has a flat dispersion, we  replace the $\delta$ functions with a Lorentzian ${\cal L}_\gamma(E)$ = $\gamma/\pi(E^2+\gamma^2)$, where $\gamma$ is the broadening of Landau levels, which can be again attributed to disorder.
Using again Eq.\,(\ref{Eq:phonon_rate}), we find the SO phonon scattering rate to be
\begin{align}
\label{Eq:SOrate}
\Gamma^{SO}_{n_i \to n_f} &= \frac{1}{2} e^2 \omega_{SO} \left( \frac{1}{\kappa_\infty^{sub}} - \frac{1}{\kappa_0^{sub}} \right) {\cal L}_\gamma (\epsilon_{n_i} - \epsilon_{n_f} - \hbar\omega_{SO}) \nonumber \\
&\times \int_{0}^\infty d q \dfrac{q^2}{(q- \frac{2\pi e^2}{\kappa_0 A} \Pi^0(q,\omega))^2} \int_0^{2\pi} d\theta |G_{n_f n_i}(q \sin\theta, q \cos\theta)|^2 ~,
\end{align}
where the screening effect is included, and $\omega$ = $(\epsilon_{n_i} - \epsilon_{n_f})/\hbar$.

\section{Landau level populations under optical pumping}

After the expressions for the optical transition rates and the scattering rates due to SO phonon emission and Auger processes have been found, we can write the density matrix equations with adiabatically eliminated optical polarizations to arrive at the set of rate equations for the filling factors of the Landau levels: 
\begin{align}
\frac{d}{d t} f_{n_a} = \left. \frac{d}{d t} f_{n_a} \right|^{op} + \left. \frac{d}{d t} f_{n_a} \right|^{SO} + \left. \frac{d}{d t} f_{n_a} \right|^{Auger} ~,
\end{align}
where
\begin{align}
\left. \frac{d}{d t} f_{n_a} \right|^{op} 
&= - \sum_{n_b} \Gamma^{op}_{n_a n_b} (f_{n_a}-f_{n_b})~, 
\\
\left. \frac{d}{d t} f_{n_a} \right|^{SO} &= - \sum_{n_b} \Gamma^{SO}_{n_a \to n_b} f_{n_a}(1-f_{n_b}) + \sum_{n_b} \Gamma^{SO}_{n_b \to n_a} f_{n_b}(1-f_{n_a}) ~, 
\end{align}
and 
\begin{align}
&\left. \frac{d}{d t} f_{n_a} \right|^{Auger} = \sum_{n_b} t_{n_a,n_b} \nonumber \\
 & \times \sum_{n_c} \sum_{n_d \ge n_c} \Gamma_{n_a n_b \leftrightarrow n_c n_d} \left( -f_{n_a} f_{n_b} (1-f_{n_c}) (1-f_{n_d}) + f_{n_c} f_{n_d} (1-f_{n_a}) (1-f_{n_b}) \right)~,
\end{align}
where $t_{n_a,n_b}$ = 2 if $n_b$ = $n_a$, and 1 otherwise.

Using these rate equations, we can simulate the dynamics of the graphene system for an arbitrary optical excitation. Note that the system is highly nonlinear, firstly because of the state filling and secondly, because the matrix elements depend on the dynamic screening, which depends in turn on the instantaneous distribution of electrons in Landau levels. Therefore, time dependent simulations are time consuming. Here we present the steady state results for the continuous-wave optical pumping.

\section{Results and discussion}

For a GaAs substrate, the SO phonon energy is 36\,meV, which requires the magnetic field to be around 1\,T. In the simulations, broadenings of all transitions are set to be 5\,meV, and $T_2$ is 0.1\,ps. Also, we consider intrinsic (undoped) graphene as an example, so without pumping the $n = 0$ LL is half-filled, all LLs below are fully filled, and all LLs above are empty.  We define the gain between $n = -1$ and $-2$ as the left-hand side of Eq.~(\ref{Eq:gain_requirement}): $g_{\mbox{\tiny -1,-2}}$ = $- ({2\pi \omega} / \sqrt{\kappa}c) {\rm Im}[\chi(\omega_{\mbox{\tiny -1,-2}})]$. 
To minimize the absorption of the THz field by the polar substrate, we would like to reduce its thickness to a few $\mu$m to be much smaller than the wavelength of the THz field but at the same time, thick enough to be considered bulk for SO phonon scattering. The gain is maximized when there is air outside the active layer so that $\kappa$ = 1.

\begin{figure}[htbp]
\includegraphics[width=0.5\textwidth]{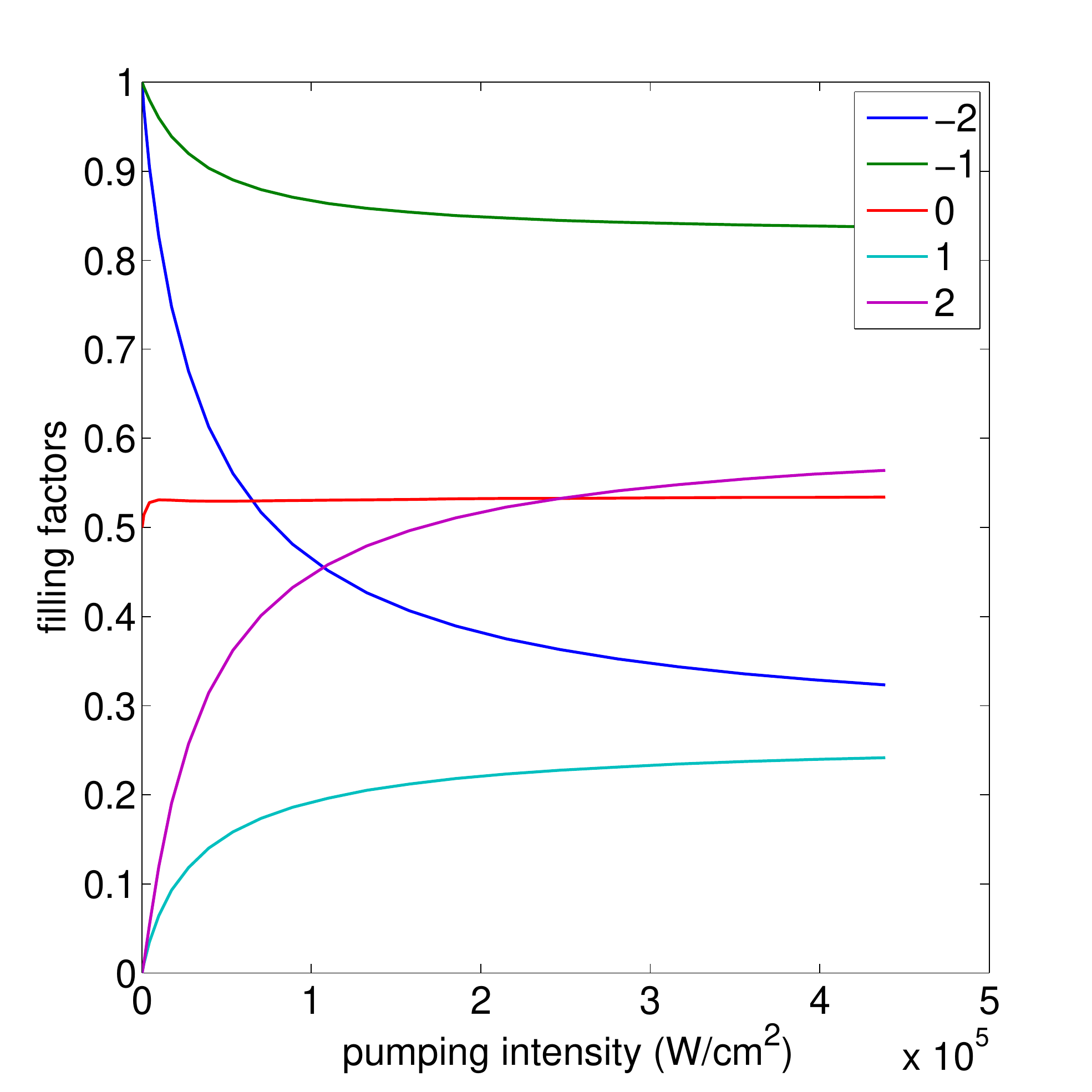}
\caption{Dependence of the steady state filling factors on the pumping intensity.\label{Fig:ff_Ipump}}
\end{figure}

\begin{figure}[htbp]
\includegraphics[width=0.5\textwidth]{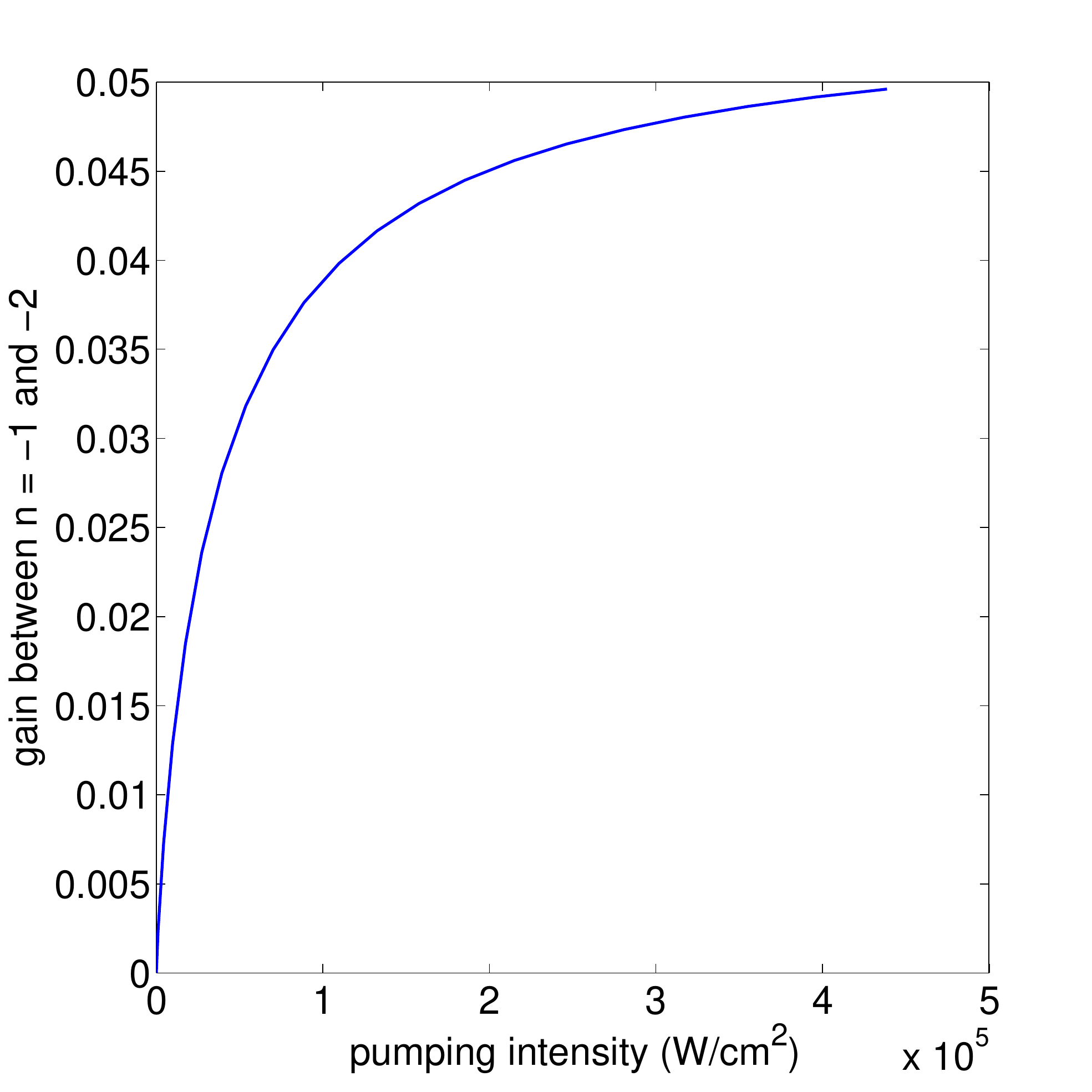}
\caption{Dependence of the gain between n = -1 and -2 LLs per graphene monolayer on the pumping intensity.\label{Fig:gain_Ipump}}
\end{figure}

The dependence of the steady state filling factors and gain per graphene monolayer on the pump intensity are shown in Fig.\,\ref{Fig:ff_Ipump} and Fig.\,\ref{Fig:gain_Ipump}.

As can be seen from the figures, one can achieve a significant steady-state population inversion between states $n = -1$ and $-2$ and the gain value of about 0.05 per monolayer of graphene. One can scale the gain up by stacking many graphene monolayers. For comparison, the right-hand side of Eq.~(\ref{Eq:gain_requirement}) which describes mirror losses is equal to 0.025 when the reflectivities $r_1 = r_2 = 0.95$, which is easily achievable. The closest allowed transition at the $\hat{l}_\ominus$ polarization is from $n = -3$ to $ n = -2$ LLs. It is detuned from the laser transition frequency by about 4 meV in a magnetic field of 1 T. Therefore its contribution to losses is lower than the gain. Since the electron motion is quantized, there are no other losses in graphene associated with free carriers. The undoped GaAs is a popular material for the nonlinear THz generation and its THz losses are rather low, especially since the polar substrate can be thinned down to a few $\mu$m. Therefore, one can operate in the desirable regime where the losses are dominated by mirror losses.  

For surface states in 3D topological insulators such as Bi$_2$Se$_3$ or  Bi$_2$Te$_3$, the Fermi velocity has a similar value but there is no spin or valley degeneracy. Therefore, a similar laser scheme with a thin film of Bi$_2$Se$_3$ (i.e. two surfaces) placed on a polar substrate will give about two times smaller gain. Additional free-carrier THz losses may exist in this case due to unintentional doping of the bulk Bi$_2$Se$_3$.

One can also see in Fig.\,\ref{Fig:ff_Ipump} that the population inversion exists also between states $n = 2$ and 1, albeit at a two times lower level. This seems unexpected, given that the optical pumping brings carriers only to state 1. However, a closer look at the rate equations shows that the population inversion between levels 1 and 2 is a consequence of a strongly non-equilibrium carrier distribution below the Fermi level created by the optical pumping, namely the population inversion between states $-2$ and $-1$. Indeed, when $f_{-1} > f_{-2}$, the Auger scattering rate from states $(1,-1)$ to states $(2,-2)$ is greater than the scattering rate in the opposite direction. This creates the population inversion $f_{2} > f_{1}$ and the gain for the $\hat{l}_\oplus$ polarization, which is about two times smaller than the $\hat{l}_\ominus$ gain. 

In conclusion, we show the feasibility  of the Landau level THz laser in a magnetized graphene.  Despite ultrafast Auger relaxation, steady-state operation of the laser under continuous wave optical pumping is possible by  utilizing surface or interface phonon relaxation. The scheme is scalable to thin films of 3D topological insulators such as Bi$_2$Se$_3$ or Bi$_2$Te$_3$.

\section*{Acknowledgments}
This work has been supported by NSF Grants OISE-0968405 and EEC-0540832, and by the AFOSR grant FA9550-14-1-0376. M. D. Tokman acknowledges support by the Russian Foundation for Basic Research Grants No. 13-02-00376 and No. 14-22-02034.

\bibliography{graphene_laser_ref}

\end{document}